\RequirePackage{lineno}
\documentclass[final,5p,times,twocolumn]{elsarticle}

\usepackage{subfigure}
\usepackage{graphicx}
\usepackage{dcolumn}
\usepackage{multirow}
\usepackage{amssymb}

\usepackage{hyperref} 

\usepackage{lineno}

\usepackage{atlasphysics}

\def\LQ{\mathrm{LQ}}
\def\LQLQ{\LQ\,\overline{\LQ}}
\def\eejj{ee jj}
\def\enjj{e\nu jj}

\begin{document}

\title{\vspace{-0.5cm} \flushright{\normalsize{CERN-PH-EP-2011-202}} \\ 
\center{Search for first generation scalar leptoquarks in $pp$ collisions \\ at $\sqrt{s}=7$~TeV with the ATLAS detector}} 

\journal{Physics Letters B}

\author{The ATLAS Collaboration}

\begin{abstract}
We report a search for first generation scalar leptoquarks using 1.03
fb$^{-1}$ of proton-proton collisions data produced by the Large Hadron
Collider at $\sqrt{s}=7$~TeV and recorded by the ATLAS experiment. Leptoquarks are 
sought via their decay into an electron or neutrino and a quark, producing events with two oppositely 
charged electrons and at least two jets, or events with an electron, missing transverse momentum and 
at least two jets. Control data samples are used to validate background predictions
from Monte Carlo simulation. In the signal region, the observed event yields are consistent with the background expectations.
We exclude at 95\% confidence level the production of first generation scalar leptoquark with masses
$m_{LQ}<$ 660 (607)~GeV when assuming the branching fraction of a leptoquark to a
charged lepton is equal to 1.0 (0.5).
\end{abstract}

\maketitle

\section{Introduction}

Similarities between leptons and quarks in the Standard Model (SM) 
suggest that they might be a part of some symmetry at energy scales above the electroweak 
symmetry breaking scale. In this type of symmetry, transitions between leptons and quarks, mediated 
by a new type of gauge boson, a leptoquark (LQ), may occur. LQs are putative color-triplet 
bosons with spin 0 or 1, and fractional electric charge~ \cite{b-lqtheo}. They are predicted in 
many extensions of the SM, such as Grand Unification models, and possess both quark and lepton 
quantum numbers. The Yukawa coupling $\lambda_{LQ-l-q}$ of a leptoquark to a lepton and a quark, 
and the branching ratio ($\beta$) to a charged lepton, are model dependent. In $pp$ collisions, 
if $\lambda_{LQ-l-q}$ is of the order of the electroweak coupling strength, leptoquarks are 
predominantly produced in pairs via the strong interaction. At the LHC, the pair production cross 
section is dominated by gluon fusion for LQ masses $m_{LQ}\lesssim 1$~TeV, whereas at higher masses 
it is dominated by quark-antiquark annihilation. Under these assumptions, the production rate 
for scalar LQs depends only on the known QCD coupling constant and the unknown LQ mass, 
and has been calculated at up to next-to-leading order. It is usually assumed that leptoquarks 
only couple to one generation of SM isospin multiplet to accommodate experimental constraints on 
flavor-changing neutral currents, and lepton and baryon number violation~\cite{b-kovalenko}. 
Consequently, they are  classified as first-, 
second-, or third-generation according to the fermion generation to which they couple~\cite{buchm}. 
Lower mass limits on the first generation LQs already exist from searches of LQ produced in pairs at the 
LHC~\cite{b-cms-e,b-lq10pub}, Tevatron~\cite{b-d0-e} and LEP~\cite{b-opal}. Limits on single LQ
production come from HERA~\cite{b-hera} and other experiments~\cite{b-pdg}.

In this Letter we present updated results on a search for the pair production of first generation 
scalar leptoquarks in $pp$ collisions at $\sqrt s=7$~TeV. The search is performed with a dataset 
corresponding to an integrated luminosity of  $1.030\pm0.035$~fb$^{-1}$~\cite{b-lumi} 
of data collected by the ATLAS detector at the LHC from March 2011 to July 2011. 
We search for leptoquarks in two different final states. In the first one both LQs decay into an 
electron and a quark, while in the second final state one of the LQs decays into an electron and a quark and 
the other LQ decays into an electron-neutrino and a quark. These result in two different 
experimental signatures. One such signature is the production of two electrons and two jets
and the other one comprises one electron, two jets, 
and missing transverse momentum (the magnitude of which is denoted as $\MET$).
The results from the two final states are combined and presented in the $m_{LQ}$ versus $\beta$ plane,
where $\beta$ is the branching ratio for a single LQ to decay into a charged lepton and a quark.

\section{The ATLAS detector}

The ATLAS detector~\cite{atlas} is a general-purpose particle detector with cylindrical geometry~\footnote{ATLAS uses a right-handed coordinate system with its origin at the nominal interaction point in the center of the detector and $z$ axis coinciding with the axis of the beam pipe. The $x$ axis points from the interaction point to the center of the LHC ring, and the $y$ axis points upward. Cylindrical coordinates $(r,\phi)$  are used in the transverse plane, $\phi$ being the azimuthal angle around the beam pipe. The pseudorapidity is defined in terms of the polar angle $\theta$ as $\eta=-$ln tan$(\theta/2)$.}, 
which consists of several subdetectors surrounding the interaction point, and providing nearly 4$\pi$ coverage
in solid angle. The location of the interaction point and momenta of charged particles 
are determined by the multi-layer silicon pixel and strip detectors covering $|\eta|<$~2.5 in pseudorapidity $\eta$, and a transition radiation tracker extending to $|\eta|<$~2.0, which are inside a superconducting solenoid 
producing a field of 2~T. The tracking system is surrounded by a high-granularity liquid-argon (LAr) sampling 
electromagnetic calorimeter with coverage up to $|\eta|<3.2$. An iron-scintillator tile hadronic 
calorimeter provides coverage in the range $|\eta|<1.7$. In the end-cap and forward regions LAr calorimeters 
provide both electromagnetic and hadronic measurements and cover the region $1.5<|\eta|<4.9$. 
The muon spectrometer, consisting of precision tracking detectors and superconducting toroids, 
is located outside the calorimeters.

We perform the search in the data sample selected by a three-level trigger requiring at least one 
high transverse energy ($E_{\rm T}$) electron. The trigger is fully efficient for electrons with 
$E_{\rm T}>30$ GeV, as measured in an inclusive $Z\to ee$ control sample~\cite{b-etrigger}.

\section{Simulated samples}

Samples of Monte Carlo (MC) events are used to devise selection criteria and validate background
predictions. Background and signal samples are processed through the full ATLAS detector
simulation based on GEANT4~\cite{b-geant4}, followed by the same reconstruction algorithms
as used for collision data. The effects from in-time and
out-of-time proton-proton collisions are included in the MC simulation. 
In the simulated samples, an event weight is applied to the average number of additional 
proton-proton collisions occurring in the same bunch crossing (event pile-up), to ensure that 
the number of interactions per bunch crossing, amounting to an average of 6, is well modeled.   

The dominant backgrounds to the leptoquark signal include $W$ and $Z$ 
boson production in association with one or more jets, single and pair production of top quarks, 
QCD multi-jet (MJ) and diboson processes. The ALPGEN~\cite{b-alpgen} generator is 
used for the simulation of the $W, Z$ boson  production in association with $n$ partons. This program is 
interfaced to HERWIG~\cite{b-herwig} and JIMMY~\cite{b-jimmy} to model parton showers and multiple parton 
interactions, respectively. The MLM~\cite{b-alpgen} jet-parton matching scheme is used to form inclusive 
$W/Z+$jets MC samples. MC@NLO~\cite{b-mcnlo} is used to estimate single and pair production of top quarks.
Diboson events are generated using HERWIG, and scaled to next-to-leading (NLO) cross section predictions~\cite{b-mcnlo,b-mcfm}.

Signal LQ samples are produced with PYTHIA~\cite{b-pythia} and normalized with NLO cross-sections 
determined from Ref.~\cite{b-kramer} using CTEQ6.6~\cite{b-cteq66} parton distribution functions.

\section{Object identification}
 \label{sec-objsel}

This search is based on selecting events with a high $E_{\rm T}$ electron, two high $p_{\rm T}$ jets, and an additional 
electron or large $\MET$. Electron candidates are reconstructed as energy deposits in the electromagnetic 
calorimeter. Electrons are required to have a shower profile consistent with that expected for this particle, and
to have a track pointing to the energy deposit in the calorimeter. The pattern of the energy deposits on 
the first layer of the EM calorimeter is used to reject hadrons, while contamination from photon conversions 
is reduced by requiring a hit in the first layer of the pixel detector~\cite{b-eperf}. In addition to these 
criteria, we require electrons to have a transverse energy $E_{\rm T}>30$~GeV and fall within a well instrumented region of the detector. 
Further rejection against hadrons is achieved by requiring the electron candidates to be isolated 
from additional energy deposits in the calorimeter by requiring that $E^{0.2}_{\rm T}/E_{\rm T}<0.1$, 
where $E^{0.2}_{\rm T}$ is the transverse energy in a cone of radius $\Delta R= \sqrt{(\Delta\eta)^2+(\Delta\phi)^2}=0.2$ centered on the electron 
track, excluding the electron contribution, and corrected for the energy from event pile-up and the 
electron energy leakage inside the cone.

Jets are defined as localized energy deposits in the calorimeter and are reconstructed using the anti-$k_t$ 
algorithm~\cite{b-akt} with a distance parameter of 0.4 and by performing a four-vector sum over calorimeter clusters. 
Reconstructed jets are corrected for the non-compensating calorimeter response, upstream material and other effects 
by using $p_{\rm T}$- and $\eta$- dependent correction factors derived from MC and validated with 
test-beam and collision data~\cite{b-jes}. We further require that jets satisfy 
$E_{\rm T}>30$ GeV, $|\eta| < 2.8$ and are separated from electrons passing the above selection within 
$\Delta R > 0.4$. Selected jets must also pass quality requirements to reject jets arising from 
electronic noise bursts, cosmic rays and beam background, originating mainly from beam-gas events and
beam-halo events~\cite{b-cleaning}.

The presence of neutrinos is inferred from the missing transverse momentum $\vec{p}_{\rm T}^{\rm miss}$ 
(and its magnitude $\MET$)~\cite{b-MET}. $\vec{p}_{\rm T}^{\rm miss}$ is defined as the negative vector sum 
of the transverse momenta of reconstructed electrons, muons and jets, as well as calorimeter clusters not 
associated to reconstructed objects.

Corrections are made to the simulated samples to ensure a good description of the
energy resolution and the trigger and reconstruction 
efficiencies. These are determined in control data samples and applied to both simulated background and signal
samples. These corrections change the total expected yields by less than 2\%.
 
\section{Event selection}
\label{sec-sel}

We define event selections to create samples with high signal and background acceptance.
Events are selected to be consistent with 
the $\LQLQ \to eeq\bar q/e\nu q\bar q$ decays. In the $\eejj$ topology we require two 
electrons and at least two jets as defined in Section~\ref{sec-objsel} and
an invariant mass of the electron pair $m_{ee}>40$~GeV. In the $\enjj$ topology,
one electron, at least two jets and $\MET>30$~GeV are required, together with
a requirement on the transverse mass of the electron and the $\vec{p}_{\rm T}^{\rm miss}$, 
$m_{\rm T}=\sqrt{2\vec{p}^e_{\rm T}\vec{p}_{\rm T}^{\rm miss}(1-\cos(\Delta\phi))}>40$~GeV, where $\Delta \phi$ 
is the angle between the electron $p_{\rm T}$ and $\vec{p}_{\rm T}^{\rm miss}$. In addition, we require that 
$\Delta\phi({\rm jet},\vec{p}_{\rm T}^{\rm miss})>4.5\times(1-\MET/45{\rm GeV})$ in the $\enjj$ channel 
for events with $\MET<45$~GeV to reduce residual contamination 
from MJ events. Events with additional identified electrons as defined in Section~\ref{sec-objsel} 
or muons with $p_{\rm T}>30$~GeV and $|\eta|<2.4$ are rejected. 

After all the selection criteria are applied the signal acceptance is of 70\% for a LQ signal
of $m_{LQ}=600$~GeV for both channels, but the sample is still dominated by background 
events.

\section{Background determination}
\label{sec-bkg}

The MJ background estimate is derived directly from data,   
whereas MC samples are used to predict the other backgrounds.   
We verify the shape of the $V$+jets $(V=W^{\pm}, Z)$ and top quark background   
prediction using control regions, which are defined to enhance   
either the $V+$jets or the top quark production contribution,   
while keeping a negligible LQ signal contamination. These control   
regions are also used to derive the final normalization of   
the $V$+jets and top quark backgrounds.

The $V$+jets and top quark control regions are defined by applying additional selection criteria on 
$m_{ee}$ and $m_{\rm T}$ to the selected sample.
The remaining signal contamination is reduced by applying an upper threshold to the summed
transverse momentum in the event, $S_{\rm T}$, defined as the scalar sum of the $p_{\rm T}$ of the two 
leading jets and the transverse energy of the
two electrons in the $\eejj$ channel. In the $S_{\rm T}$ definition in the $\enjj$ channel, 
the second electron $E_{\rm T}$ is substituted by the $\MET$.

In the $\eejj$ topology we define two control regions {\bf (i)} $Z+$jets: formed by events with at least
two jets and in which
the two electrons are required to have an invariant mass within a $Z$ mass window $\mbox 81<m_{ee}<101$~GeV,
and {\bf (ii)} $t\bar t$: events with at least two jets and exactly one electron and one muon~\cite{b-mcp},
defined as in Section~\ref{sec-objsel}.
In the $\enjj$ topology we define three control regions {\bf (iii)} $W+2$ jets: events with exactly two jets,
an electron and $\MET$ such that the transverse mass of the electron and the $\MET$ is in the region
of the $W$ Jacobian peak, $\mbox 40 <m_{\rm T}<120$~GeV, and a $S_{\rm T}<225$~GeV requirement to limit the 
presence of signal events, {\bf (iv)} $W+3$ jets: as in {\bf (iii)} but with three or more jets,
and {\bf (v)} $\ttbar$: events with at least 4 jets, where the thresholds on the first and second jets are 
raised to 50~GeV and 40~GeV, respectively.

To estimate the MJ background, we perform fits to the $m_{ee}$ distribution in the $\eejj$ channel, 
and to the $\MET$ distribution in the $\enjj$ channel. In these fits, the relative fraction 
of the MJ background is a free parameter. 
Templates for the MJ background distributions are derived from MJ enhanced samples, which are   
formed using electron candidates passing relaxed selection requirements but failing the nominal 
electron identification criteria described in Section~\ref{sec-objsel}. The MJ enhanced samples are   
corrected to remove the residual contamination from real electrons. In the $\eejj$ channel, the 
fits are applied to the sample selected following the criteria of Section~\ref{sec-sel}, as well as to   
control regions {\bf (i)} and {\bf (ii)}, and the W+jets background is estimated together with the MJ background. 
In the $\enjj$ channel, the fits are applied to the selected sample as well as to control regions {\bf (iii-v)}.

We observe 5615 data events in the $\eejj$ channel and 76855
data events in the $\enjj$ channel, with SM expectations of $5600\pm1000$ and $74000\pm11000$, respectively.
For $m_{LQ}=600$~GeV, we expect $7.5\pm 0.5$ signal events in the $\eejj$ channel and $4.5\pm0.2$ signal
events in the $\enjj$ channel. The aforementioned uncertainties fully account for (the dominant) 
systematic and statistical uncertainties.

\begin{figure*}[htbp!]
\begin{center}
    \subfigure[]{\includegraphics[width= 0.49\textwidth]{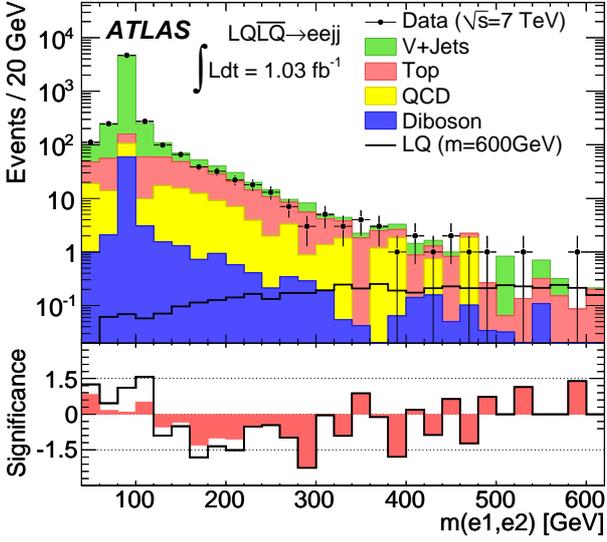}\label{f-mee}}
    \subfigure[]{\includegraphics[width= 0.49\textwidth]{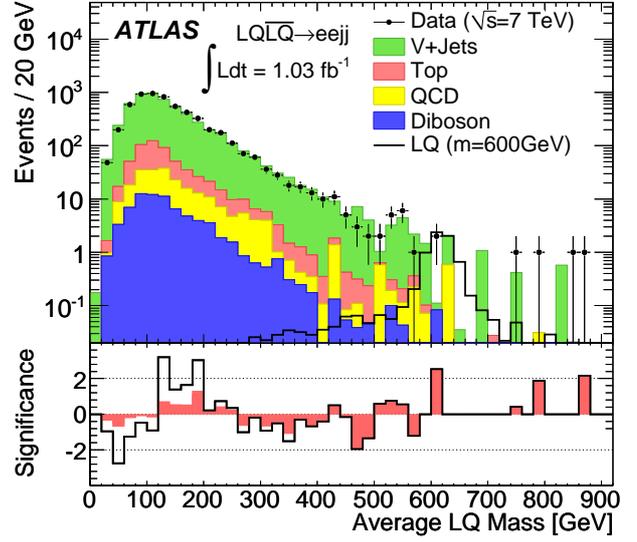}\label{f-lqm}}
    \subfigure[]{\includegraphics[width= 0.49\textwidth]{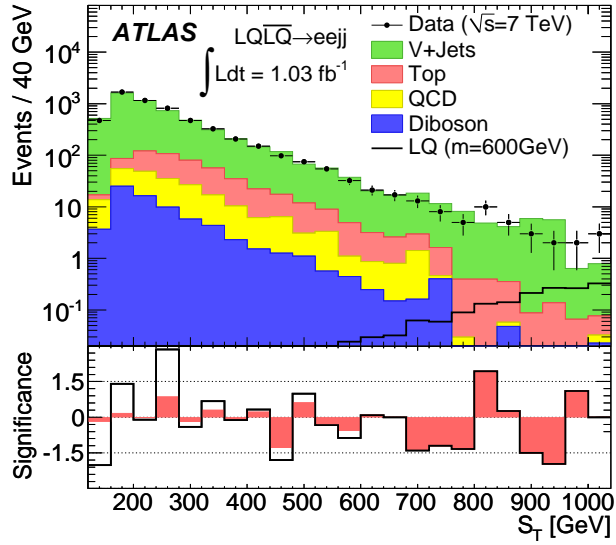}\label{f-st}}
\end{center}
    \caption{Data and SM background comparisons of the input $LLR$ variables for
    the $\eejj$ channel. \subref{f-mee} Invariant mass of the two electrons in the event, 
    \subref{f-lqm} Average LQ mass resulting from the best (electron, jet) combinations in each event, 
    and \subref{f-st} $S_T$. The stacked distributions show the various background
    contributions, and data are indicated by the points with error bars. The 600~GeV LQ signal is
    also shown for $\beta=1.0$. The solid line (band) in the lower plots shows the Gaussian statistical
    (statistical + systematic) significance of the difference between data and the prediction.\label{f-eejjInputs}} 
\end{figure*}

\begin{figure*}[!htbp]
\centering
  \subfigure[]{\includegraphics[width= 0.49\textwidth]{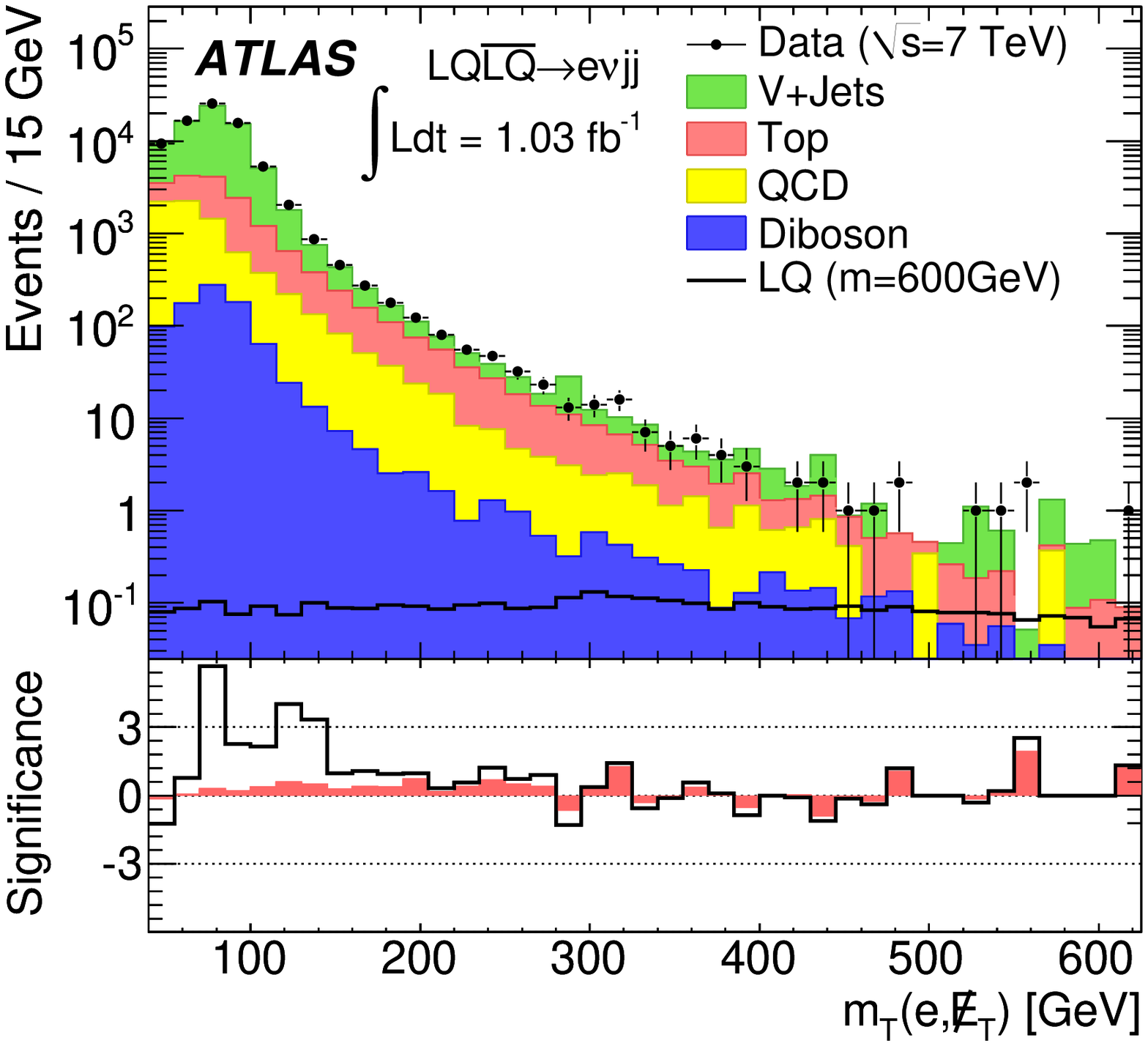}\label{f-mt}}
  \subfigure[]{\includegraphics[width= 0.49\textwidth]{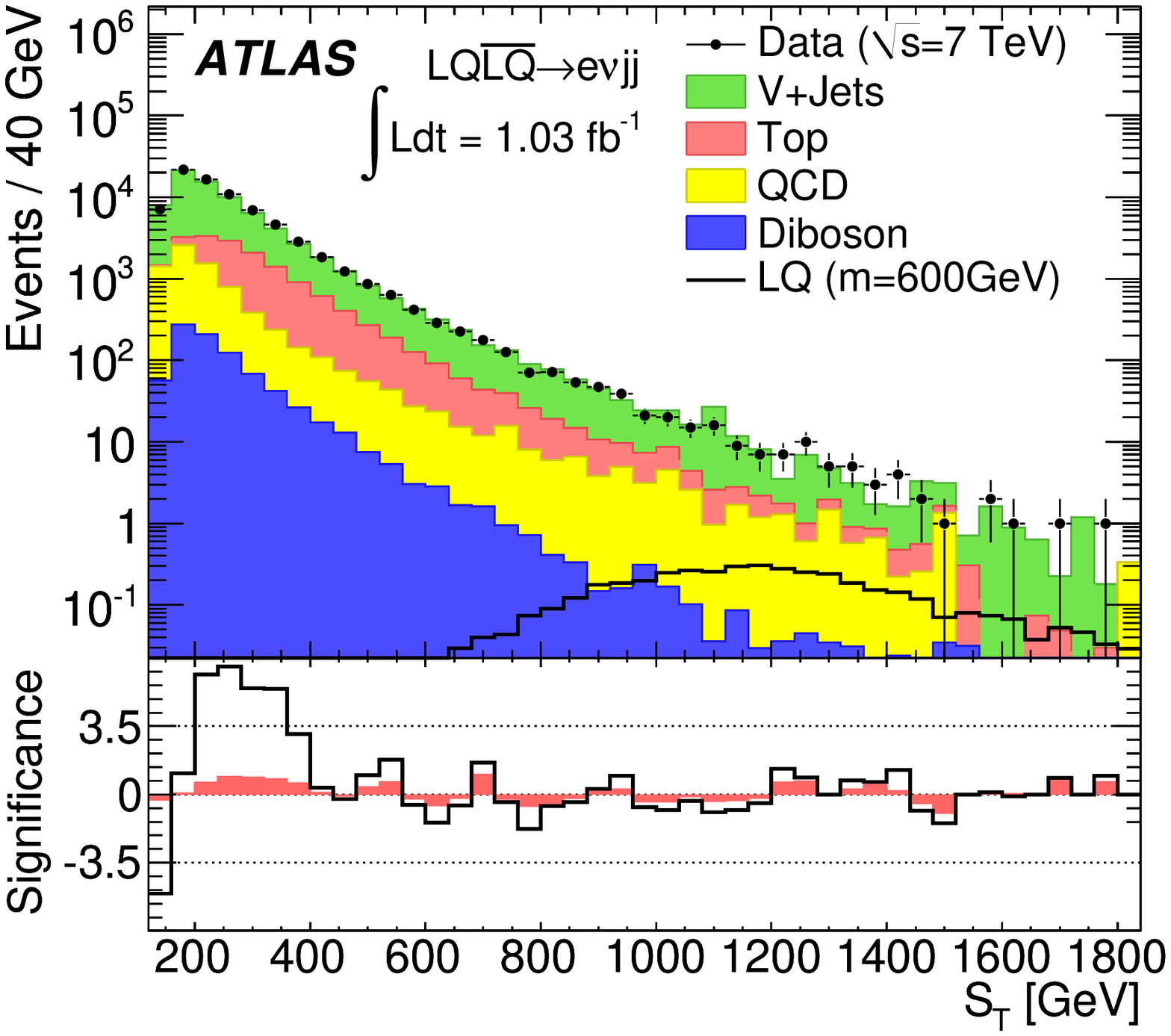}\label{f-st-evjj}}
  \subfigure[]{\includegraphics[width= 0.49\textwidth]{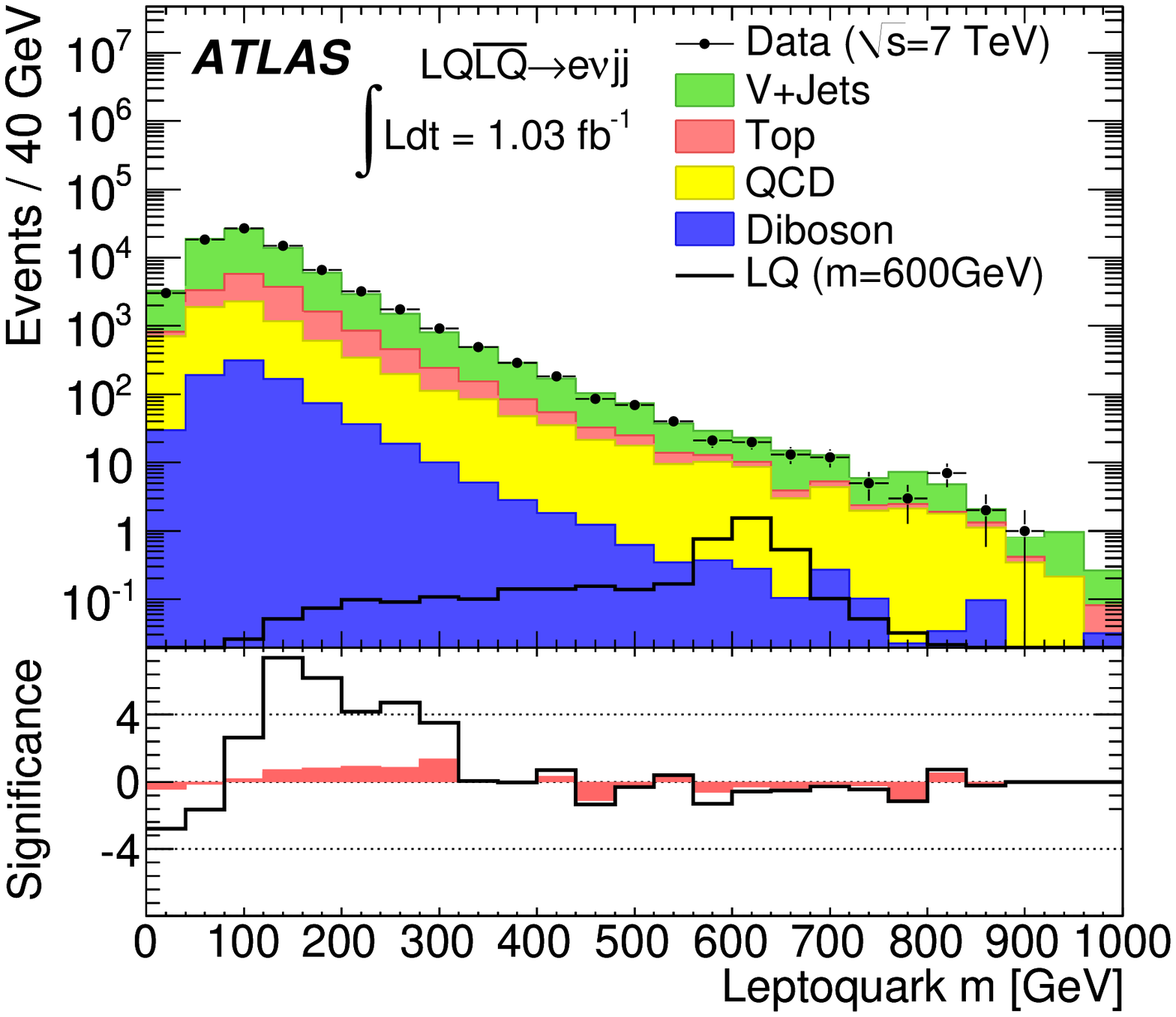}\label{f-lqm-evjj}}
  \subfigure[]{\includegraphics[width= 0.49\textwidth]{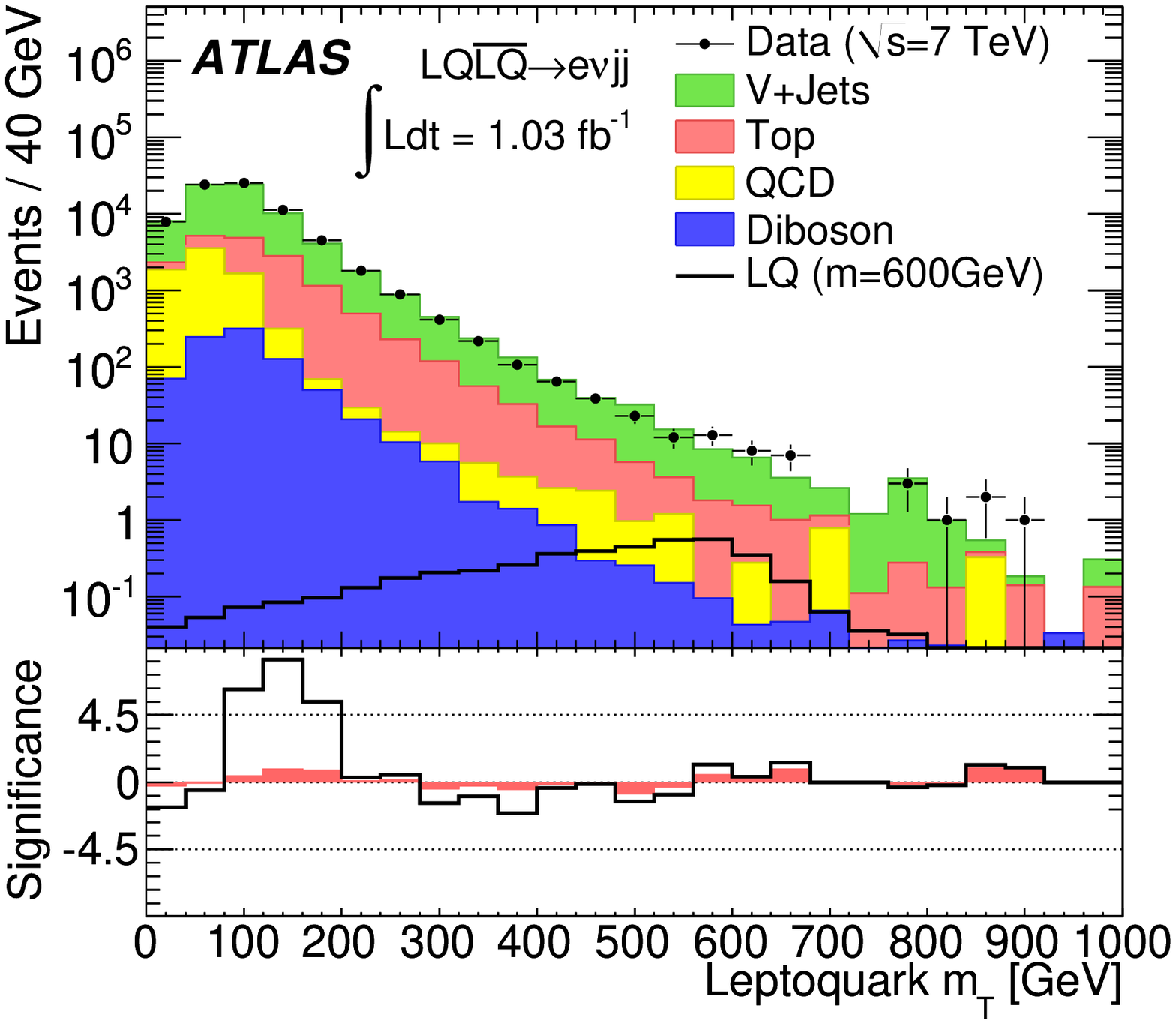}\label{f-lqmt}}
  \caption{Data and SM background comparisons of the input $LLR$ variables for
  the $\enjj$ channel. \subref{f-mt} Transverse mass of the electron and the $\MET$ in the event,
  \subref{f-st-evjj} $S_T$, \subref{f-lqm-evjj} LQ mass, and \subref{f-lqmt} LQ transverse masses.
  The stacked distributions show the various background
  contributions, and data are indicated by the points with error bars. The 600~GeV LQ signal is
  also shown for $\beta=0.5$. The solid line (band) in the lower plots shows the Gaussian statistical
    (statistical + systematic) significance of the difference between data and the prediction.\label{f-evjjInputs}}
\end{figure*}

\section{Likelihood analysis}

We use a likelihood ratio method to separate signal and SM background. The likelihoods are constructed separately 
for background ($L_B$) and signal ($L_S$) hypotheses from a set of discriminating variables as follows: 
$L_B \equiv \prod b_i(x_j)$, $L_S \equiv \prod s_i(x_j)$, where $b_i$, $s_i$ are the probabilities of the
$i$--th input variable from the normalized summed background and signal distributions respectively, and $x_j$ is 
the value of that variable for the $j$--th event in a given sample. Separate $L_S$ distributions are created 
for several signal mass points, allowing mass-dependent optimization. Using the aforementioned 
quantities, a likelihood ratio is defined as $LLR={\rm log}(L_S/L_B)$ and is used as the final variable 
to determine whether or not there is a LQ signal present in our data. 

The following discriminating variables, selected to give the best separation between signal and 
background, are used. For the $\eejj$ channel, we use $m_{ee}$, 
$S_{\rm T}=E_{\rm T}^{\rm{e1}}+E_{\rm T}^{\rm{e2}}+p_{\rm T}^{\rm{jet1}}+p_{\rm T}^{\rm{jet2}}$ 
and the average invariant LQ mass $\bar m_{LQ}$. For the $\enjj$ topology, we use $m_{\rm T}(e,\MET)$, 
$S_{\rm T}$, the transverse LQ mass $m_{\rm T}^{LQ}(\rm {jet},\MET)$ and the invariant LQ mass 
$m_{LQ}(\rm {e},\rm{jet})$. To obtain the LQ masses, we calculate the invariant mass of the 
electron-jet system and the transverse mass of the $\MET$-jet system. Since the LQs are produced in pairs,
there are two possible mass combinations for the electron-jet and $\MET$-jet pairs, and the combination 
giving the smallest mass difference is used. In the $eejj$ channel, two possible electron-jet
combinations arise from this procedure, and we take their average
$\bar m_{LQ}$ for the analysis. The discriminating variables are shown in 
Figs.~\ref{f-eejjInputs} and~\ref{f-evjjInputs} for the $\eejj$ and the $\enjj$ channels, respectively.

\section{Systematic uncertainties}

Systematic uncertainties affect both background normalizations and shapes of the input 
distributions into the $LLR$. We consider systematic uncertainties from a variety of sources.
These are described as follows. 

The jet energy scale (JES) and resolution (JER) uncertainties are considered independently, 
and applied by varying the JES (JER) within its uncertainty of 4\% to 6.5\% (14\%) depending 
on the jet $p_{\rm T}$ and $\eta$~\cite{b-jes-unc,b-jer-unc} for all simulated events. These 
variations are also propagated to the $\MET$ in the $\enjj$ channel.  The resulting 
uncertainties for the $m_{LQ}=600$~GeV signal and background are 5\% (8\%) and 11\% for the
$\eejj$ ($\enjj$) final state.

Systematic uncertainties on the electron energy scale ($1.6\%$) and resolution ($0.6\%$), and
on the electron trigger, reconstruction and identification efficiencies are derived by varying 
the selection criteria defining the Drell-Yan control sample used for the various 
measurements~\cite{b-etrigger}. In addition, a 1\% uncertainty is included to account for the efficiency of the 
isolation requirement. They lead to total signal and background yield uncertainties of 
8\% and 5\% (3.5\%), respectively, for the $\eejj$ ($\enjj$) channel and for a signal of mass
$m_{LQ}=600$~GeV.

The systematic uncertainty for the production model of $V$+jets is taken to be the largest 
difference between the nominal data-driven prediction using ALPGEN and that obtained by using
SHERPA~\cite{b-sherpa}, giving an uncertainty of 1.5\% and 3\% for the 
$\eejj$ and the $\enjj$ channels, respectively. 

The systematic uncertainty for the $\ttbar$ production model are evaluated by comparing the yields 
between events generated with MC@NLO and those generated with various alternate samples. These include
samples generated with POWHEG~\cite{b-powheg}, a different top mass (170~GeV and 175~GeV instead
of the nominal value equal to 172.5~GeV), and a different amount of initial- and final-state-
radiation (ISR/FSR). The result is an uncertainty in the $\ttbar$ yield of 10\% and 15\% 
for the single electron and dielectron analyses, respectively. 

Systematic uncertainties are determined for the MJ backgrounds by comparing results from 
alternative normalizations to those from the methods described earlier. The largest variation is
taken, resulting in an uncertainty of 20\% and 28\% in the MJ normalization for the $\enjj$ and 
the $\eejj$ channels, respectively. An uncertainty of 3.7\%~\cite{b-lumi} on the integrated
luminosity is applied to both diboson and single top background yields, as well as to expected 
signal yields. 

Finally, further uncertainties on
the simulated background contributions originate from finite statistics in the MC samples
used. These range from 2\%--9\%, depending on the LQ mass under consideration. Additional signal uncertainties 
considered arise from the choice of the PDF, which results in an uncertainty on the signal 
acceptance of 1\%--8\% for LQ masses between 300~GeV and 700~GeV, and from ISR/FSR effects, 
resulting in an uncertainty of 2\% for both channels.

\begin{figure}[!htbp]
\centering
  \subfigure[]{\includegraphics[width= 0.49\textwidth]{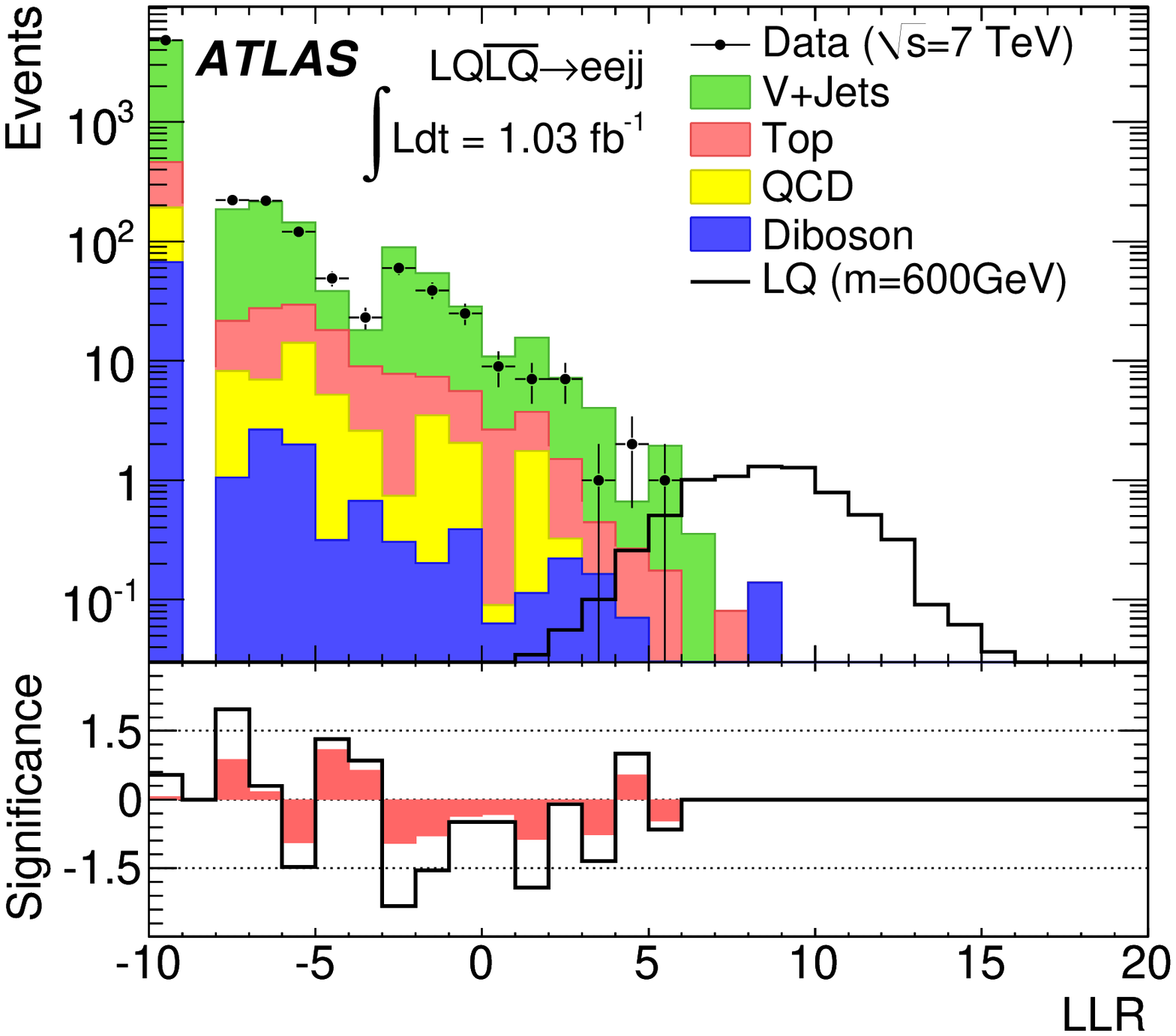}\label{f-llr-eejj}}
   \subfigure[]{\includegraphics[width= 0.49\textwidth]{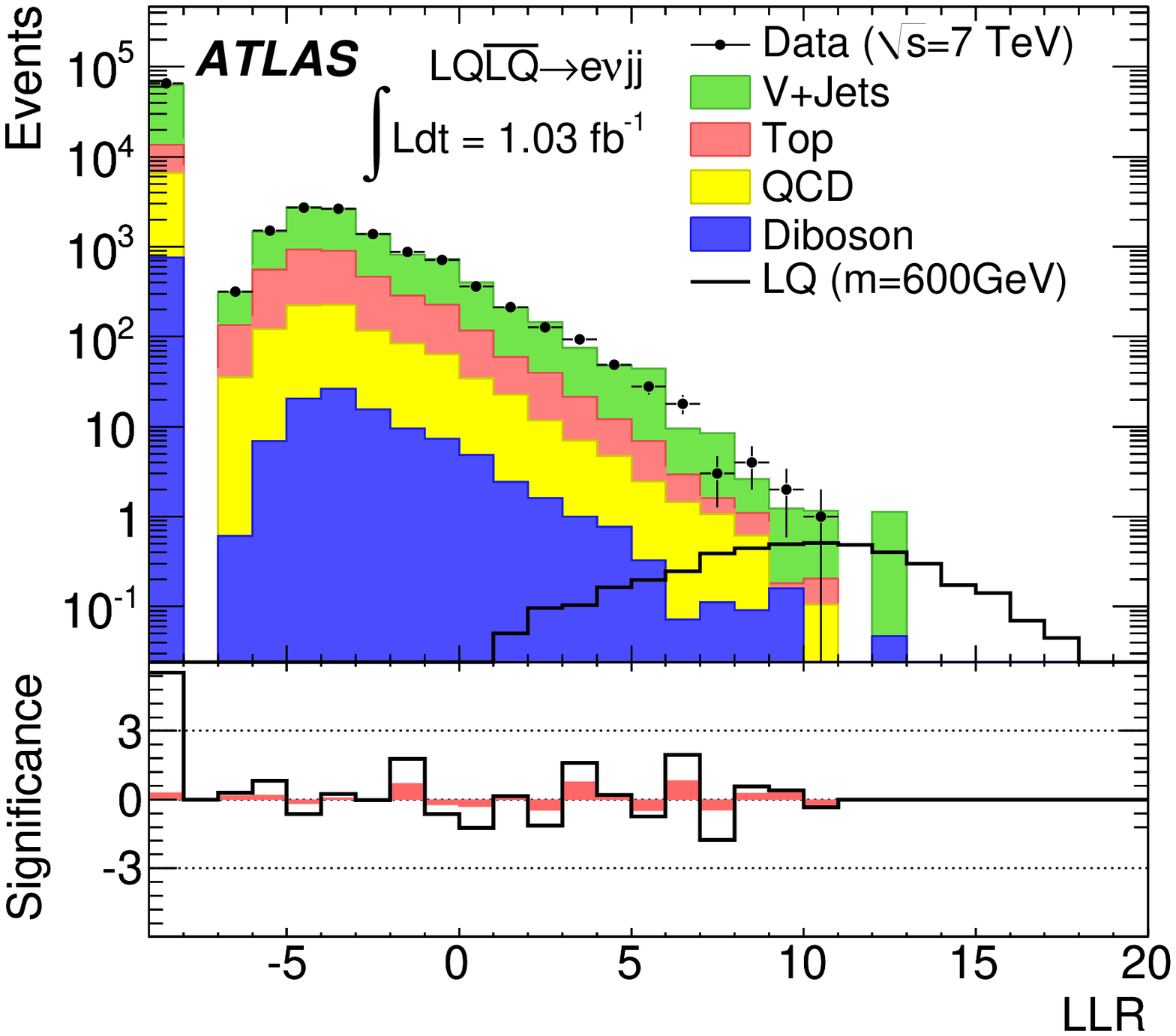}\label{f-llr-evjj}}
  \caption{$LLR$ distributions for the $\eejj$ \subref{f-llr-eejj} and for the $\enjj$ \subref{f-llr-evjj} final states.
The data are indicated with the points and the filled histograms show the SM background.
The MJ background is estimated from data, while the other background contributions
are obtained from simulated samples as described in the text. The LQ signal corresponding to a LQ mass of 600~GeV
 is indicated by a solid line, and is normalized assuming $\beta=1.0(0.5)$ in the $eejj$ ($\enjj$) channel. 
The lowest bin corresponds to background events regions of
the phase space for which no signal events are expected. The solid line (band) in the lower plots shows the Gaussian statistical
    (statistical + systematic) significance of the difference between data and the prediction.
   \label{f-llreejj}}
\end{figure}

\begin{table}
\begin{center}
\caption{The predicted and observed yields in a signal enhanced region defined by requiring $LLR>  0$ 
for both channels. Background predictions are scaled as described in Section~\ref{sec-bkg}. The 
$\eejj$ ($\enjj$) channel signal yields are computed assuming
    $\beta=1.0\,\,(0.5)$. Statistical and systematic uncertainties added in quadrature are shown.
    \label{toptsel}}
  \begin{tabular}{lcccc}\hline
  Source  & \multicolumn{2}{c}{$\eejj$ Channel} & \multicolumn{2}{c}{$\enjj$ Channel}\\
                     & 400 GeV & 600 GeV & 400 GeV & 600 GeV \\ \hline\hline
$W$+jets                   & ---              & ---              & $1500 \pm 670$ & $ 670 \pm 210$ \\
$Z$+jets & $ 98 \pm 53$ & $ 26 \pm 14$ & $  45 \pm 41$ & $  18 \pm 19$ \\
$\ttbar$& $ 15 \pm  9$ & $  4.6 \pm  2.2$ & $ 430 \pm 180$ & $ 150 \pm  38$ \\
Single $t$ & $  1.4 \pm  0.9$ & $  0.7 \pm  0.4$ & $  53 \pm  19 $ & $  23 \pm 4$ \\
 Dibosons& $  1.5 \pm  0.8$ & $  0.7 \pm  0.3$ & $  25 \pm  11$ & $  11 \pm 2$ \\
MJ& $  9.2 \pm  4.5$ & $  2.3 \pm  1.5$ & $ 170 \pm  35$ & $  75 \pm  15$ \\
\hline\hline
Total& $120 \pm 55$ & $ 34 \pm 14$ & $2200 \pm 690$ & $ 950 \pm 220$ \\
Data & $ 82$            & $  22$           & $ 2207$           & $   900 $          \\
\hline
LQ & $120 \pm 8$ & $ 7.5 \pm 0.5$ & $  69 \pm 4$  & $   4.5 \pm  0.2$  \\
\hline
\end{tabular}
\end{center}
\end{table}

\section{Results}

\begin{figure}[!htbp]
\centering
  \subfigure[]{\includegraphics[width= 0.49\textwidth]{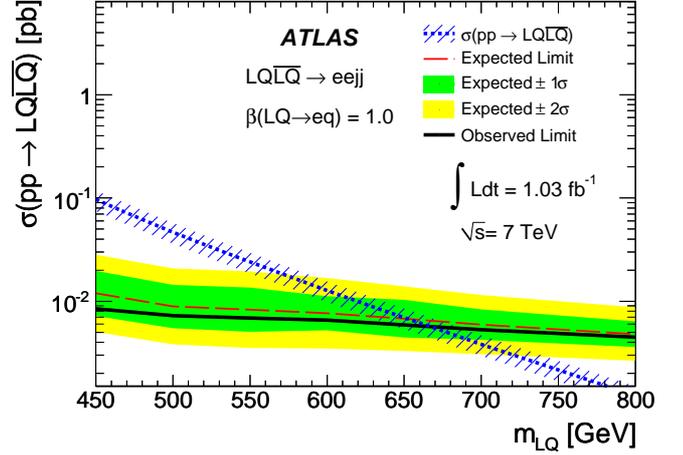}\label{f-eejj}}
  \subfigure[]{\includegraphics[width= 0.49\textwidth]{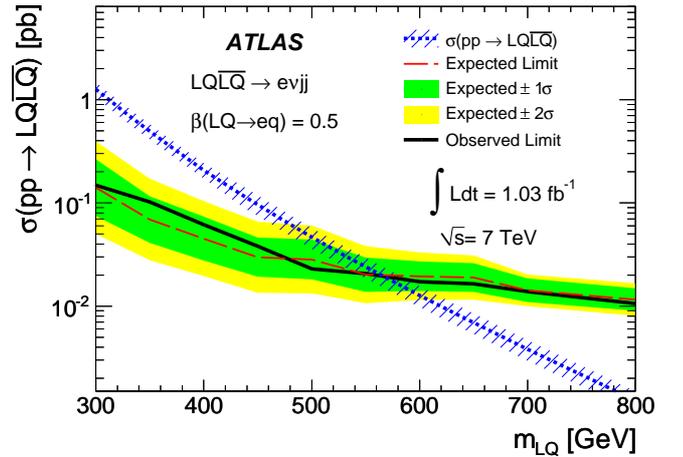}\label{f-evjj}}
    \caption{95\% CL upper limit on  the pair production cross section
    of the first generation leptoquarks for the $eejj$ channel at $\beta=$1.0 \subref{f-eejj}
    and for the $\enjj$ channel at $\beta=$0.5 \subref{f-evjj}. The solid lines indicate the individual
    observed limits, while the expected limits are indicated by the dashed lines. The theory
    prediction is indicated by the dotted line, which includes the systematic uncertainties due to
    the choice of the PDF and due to the renormalization and factorization scales.
    The dark green (light yellow) solid band contains 68\% (95\%) of possible outcomes
    from pseudo-experiments in which the yield is Poisson-fluctuated around the
    background-only expectation.
\label{f-xsec95}}
\end{figure}

\begin{figure}[!htbp]
  \subfigure[]{\includegraphics[width=0.97\columnwidth]{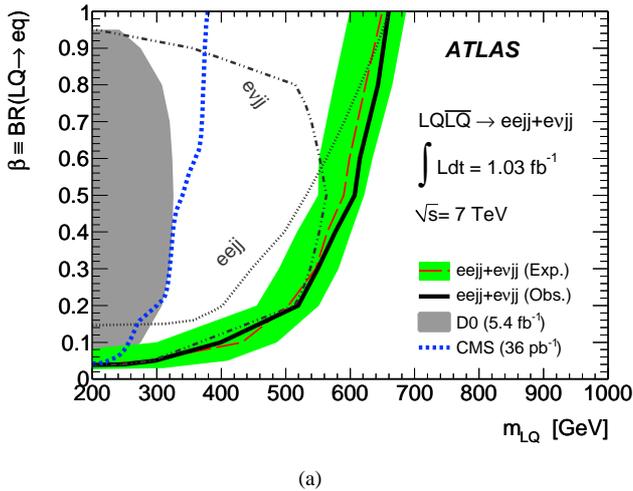}}
  \caption{95\% CL exclusion region resulting from the combination of the
  two channels shown in the $\beta$ versus leptoquark mass plane.
  The shaded
  area indicates the D0 exclusion limit~\cite{b-d0-e}, while the thick dotted line 
  indicates the CMS exclusion~\cite{b-cms-e}. 
  The dotted and dotted-dashed lines
  indicate the individual limits for the $\eejj$ and the $\enjj$, respectively.
  The combined observed limit is indicated by the solid black line. 
  The combined expected limit is indicated by the dashed line, together with
  the solid band containing 68\% of possible outcomes
    from pseudo-experiments in which the yield is Poisson-fluctuated around the
    background-only expectation.
\label{f-bvm}}
\end{figure}

The $LLR$ distributions for data, backgrounds and a LQ signal assuming $m_{LQ}=600$~GeV are shown in Fig.~\ref{f-llreejj}
for both channels. The observed and predicted event yields requiring $LLR>0$
for the major background sources, as well as the expected signal, are shown in Table~\ref{toptsel}.
We do not observe any excess of events at high $LLR$ values where 
signal is expected, indicating no evidence of scalar LQ pair production. 
Given the absence of signal we determine 95\% CL upper limits on the LQ pair--production cross sections using a 
modified frequentist $CL_s$ method based on a Poisson log-likelihood ratio statistical test~\cite{b-lcalc1,b-lcalc2}.
Systematic and statistical uncertainties are treated as nuisance parameters with a Gaussian probability 
density function, and the full $LLR$ distribution is considered. The effect of the various systematic 
uncertainties on the shape of the $LLR$ distribution are included on the calculation by integrating over a Gaussian 
distribution with standard deviation equal to the fractional change in the yield between the systematically 
adjusted distribution and the nominal case for each individual uncertainty in each bin. The 95\% CL upper bounds on 
the cross section for LQ pair production as a function of mass are shown in Fig.~\ref{f-xsec95} for both
the $\eejj$ and the $\enjj$ channels for $\beta=1.0$ and $\beta=0.5$, respectively.
The obtained cross section limits are combined, and reinterpreted as limits in the $\beta$ vs. $m_{LQ}$ 
plane as shown in Fig.~\ref{f-bvm}.

\section {Conclusions}

We report on a search for pair production of first generation scalar 
leptoquarks at ATLAS using a data sample corresponding to an integrated luminosity of 
1.03~fb$^{-1}$. No excess over SM background expectations is observed in the data in the 
signal enhanced region, and 95\% CL upper bounds on the production cross section are thus
determined. These are translated into lower observed (expected) limits on leptoquark masses 
of $m>660 (650)$~GeV and $m>607 (587)$~GeV when assuming its branching fraction to a charged lepton   
to be equal to 1.0 and 0.5, respectively. These
are the most stringent limits to date arising from direct searches for leptoquarks.

\section{Acknowledgements}



We thank CERN for the very successful operation of the LHC, as well as the
support staff from our institutions without whom ATLAS could not be
operated efficiently.

We acknowledge the support of ANPCyT, Argentina; YerPhI, Armenia; ARC,
Australia; BMWF, Austria; ANAS, Azerbaijan; SSTC, Belarus; CNPq and FAPESP,
Brazil; NSERC, NRC and CFI, Canada; CERN; CONICYT, Chile; CAS, MOST and
NSFC, China; COLCIENCIAS, Colombia; MSMT CR, MPO CR and VSC CR, Czech
Republic; DNRF, DNSRC and Lundbeck Foundation, Denmark; ARTEMIS, European
Union; IN2P3-CNRS, CEA-DSM/IRFU, France; GNAS, Georgia; BMBF, DFG, HGF, MPG
and AvH Foundation, Germany; GSRT, Greece; ISF, MINERVA, GIF, DIP and
Benoziyo Center, Israel; INFN, Italy; MEXT and JSPS, Japan; CNRST, Morocco;
FOM and NWO, Netherlands; RCN, Norway; MNiSW, Poland; GRICES and FCT,
Portugal; MERYS (MECTS), Romania; MES of Russia and ROSATOM, Russian
Federation; JINR; MSTD, Serbia; MSSR, Slovakia; ARRS and MVZT, Slovenia;
DST/NRF, South Africa; MICINN, Spain; SRC and Wallenberg Foundation,
Sweden; SER, SNSF and Cantons of Bern and Geneva, Switzerland; NSC, Taiwan;
TAEK, Turkey; STFC, the Royal Society and Leverhulme Trust, United Kingdom;
DOE and NSF, United States of America.

The crucial computing support from all WLCG partners is acknowledged
gratefully, in particular from CERN and the ATLAS Tier-1 facilities at
TRIUMF (Canada), NDGF (Denmark, Norway, Sweden), CC-IN2P3 (France),
KIT/GridKA (Germany), INFN-CNAF (Italy), NL-T1 (Netherlands), PIC (Spain),
ASGC (Taiwan), RAL (UK) and BNL (USA) and in the Tier-2 facilities
worldwide.


\clearpage 
\onecolumn
\begin{flushleft}
{\Large The ATLAS Collaboration}

\bigskip

G.~Aad$^{\rm 48}$,
B.~Abbott$^{\rm 110}$,
J.~Abdallah$^{\rm 11}$,
A.A.~Abdelalim$^{\rm 49}$,
A.~Abdesselam$^{\rm 117}$,
O.~Abdinov$^{\rm 10}$,
B.~Abi$^{\rm 111}$,
M.~Abolins$^{\rm 87}$,
O.S.~AbouZeid$^{\rm 157}$,
H.~Abramowicz$^{\rm 152}$
% ...
\bigskip

% ...
$^{10}$ Institute of Physics, Azerbaijan Academy of Sciences, Baku, Azerbaijan\\
$^{11}$ Institut de F\'isica d'Altes Energies and Departament de F\'isica de la Universitat Aut\`onoma  de Barcelona and ICREA, Barcelona, Spain\\
% ...
$^{48}$ Fakult\"{a}t f\"{u}r Mathematik und Physik, Albert-Ludwigs-Universit\"{a}t, Freiburg i.Br., Germany\\
$^{49}$ Section de Physique, Universit\'e de Gen\`eve, Geneva, Switzerland\\
% ...
$^{87}$ Department of Physics and Astronomy, Michigan State University, East Lansing MI, United States of America\\
% ...
$^{110}$ Homer L. Dodge Department of Physics and Astronomy, University of Oklahoma, Norman OK, United States of America\\
$^{111}$ Department of Physics, Oklahoma State University, Stillwater OK, United States of America\\
% ...
$^{117}$ Department of Physics, Oxford University, Oxford, United Kingdom\\
% ...
$^{152}$ Raymond and Beverly Sackler School of Physics and Astronomy, Tel Aviv University, Tel Aviv, Israel\\
% ...
$^{157}$ Department of Physics, University of Toronto, Toronto ON, Canada\\
% ...
\end{flushleft}


\end{document}